\documentclass[12pt,preprint]{aastex}
\usepackage{epsfig}

\def\degree{$^\circ$}
\def\arcs#1{$#1''$}
\def\arcsa#1#2{$#1^{\prime\prime}_{^\textrm{.}}#2$}

\def\solarmass{$M_\odot$}

\def\uJyb{$\mu$Jy beam$^{-1}$}

\def\cmc{cm$^{-3}$}
\def\cms{cm$^{-2}$}
\def\micron{$\mu$m}

\def\ra#1#2#3#4{$#1^\mathrm{h} #2^\mathrm{m} #3^\mathrm{s}_{^\textrm{.}} #4$}
\def\dec#1#2#3#4{$#1\degr #2\arcmin #3^{\prime\prime}_{^\textrm{.}}#4$}

\def\Ross{\textrm{\scriptsize R}}
\def\eff{\textrm{\scriptsize eff}}
\def\mid{\textrm{\scriptsize mid}}

\def\mHt{m_{\textrm{\scriptsize H}_2}}

\def\H2{H$_2$}
\def\N2HP{N$_2$H$^+$}

\def\NH3{NH$_3$}

\def\mH2{m_{\textrm{\scriptsize H}_2}}

\def\na{n_\mathrm{t}}
\def\rhoa{\rho_\mathrm{t}}
\def\Ro{R_\mathrm{o}}

\def\Ra{R_\mathrm{t}}
\def\hs{h_\mathrm{s}}

\def\Ta{T_\mathrm{t}}
\def\ho{h_\mathrm{o}}
\def\ha{h_\mathrm{t}}

\def\cs{c_\mathrm{s}}

\def\vp{v_\phi}

\def\kabs{\kappa_\textrm{\scriptsize abs}}
\def\ksca{\kappa_\textrm{\scriptsize sca}}

\def\iI{{\it I}}
\def\iQ{{\it Q}}
\def\iU{{\it U}}
                                                                            
\def\putfig#1#2#3{\epsfig{scale=#1,angle=#2,figure=#3}}
\def\putfiga#1#2#3{}
\def\leftblank#1{}

\begin{document}

\title{What Produces Dust Polarization in the HH 212 Protostellar Disk at
878 \micron{}: Dust Self-Scattering or Dichroic Extinction?}

\author{Chin-Fei Lee\altaffilmark{1,2}, Zhi-Yun Li\altaffilmark{3},
Haifeng Yang\altaffilmark{4}, Zhe-Yu Daniel Lin\altaffilmark{3},
Tao-Chung Ching\altaffilmark{5,6}, and Shih-Ping Lai\altaffilmark{7,1}
}

\altaffiltext{1}{Academia Sinica Institute of Astronomy and Astrophysics,
P.O. Box 23-141, Taipei 106, Taiwan; cflee@asiaa.sinica.edu.tw}
\altaffiltext{2}{Graduate Institute of Astronomy and Astrophysics, National Taiwan
   University, No.  1, Sec.  4, Roosevelt Road, Taipei 10617, Taiwan}
\altaffiltext{3}{Astronomy Department, University of Virginia, Charlottesville, VA 22904, USA}
\altaffiltext{4}{Institute for Advanced Study, Tsinghua University, Beijing, 100084, People's Republic of China}
\altaffiltext{5}{CAS Key Laboratory of FAST, National Astronomical Observatories, Chinese Academy of Sciences, People's Republic of China}
\altaffiltext{6}{National Astronomical Observatories, Chinese Academy of Sciences, A20 Datun Road, Chaoyang District, Beijing 100012, People's
Republic of China}
\altaffiltext{7}{Institute of Astronomy and Department of Physics, National
Tsing Hua University, Hsinchu, Taiwan}

\begin{abstract}

We report new dust polarization results of a nearly edge-on disk in the HH
212 protostellar system, obtained with ALMA at $\sim$ \arcsa{0}{035} (14 au)
resolution in continuum at $\lambda \sim$ 878 \micron{}.  Dust polarization
is detected within $\sim$ 44 au of the central source, where a rotationally
supported disk has formed.  The polarized emission forms V-shaped
structures opening to the east and probably west arising from the disk
surfaces and arm structures further away in the east and west that could be
due to potential spiral arms excited in the outer disk.  The polarization
orientations are mainly parallel to the minor axis of the disk, with
some in the western part tilting slightly away from the minor axis to
form a concave shape with respect to the center. This tilt
of polarization orientations is expected from dust self-scattering, e.g., by
50$-$75 \micron{} grains in a young disk.  The polarized intensity and
polarization degree both peak near the central source with a small dip at
the central source and decrease towards the edges.  These decreases of
polarized intensity and polarization degree are expected from dichroic
extinction by grains aligned by poloidal fields, but may also be
consistent with dust self-scattering if the grain size decreases toward the
edges.  It is possible that both mechanisms are needed to produce the
observed dust polarization, suggesting the presence of both grain growth and
poloidal fields in the disk.

\end{abstract}

\keywords{stars: formation --- ISM: individual: HH 212 --- ISM:
accretion and accretion disk -- ISM: magnetic fields -- polarization}

\section{Introduction}

HH 212 is a young accreting protostellar system with a highly collimated
spinning jet \citep{Zinnecker1998,Lee2017Jet} in the Class 0 phase in Orion
at a distance of $\sim$ 400 pc.  With a nearly edge-on and vertically
resolved disk \citep{Lee2017Disk} deeply embedded in a dense rotating
molecular core \citep{Wiseman2001} and an infalling-rotating flattened
envelope \citep{Lee2006,Lee2014}, it becomes a textbook case to study the
disk formation and accretion process in the earliest phase of star
formation.  The disk is rotationally supported, surrounding a protostar with
a mass of $\sim$ 0.25 \solarmass{} \citep{Codella2014,Lee2017COMs}.  Dust
polarization has been detected in the disk in submillimeter wavelength at
$\sim$ 875 \micron{} \citep{Lee2018BDisk}, and it can be due to either
dichroic extinction by grains aligned magnetically by poloidal fields or
dust self-scattering by grains with a maximum size up to 100 \micron{} in
the outer disk.  Both of these possibilities have important implications. 
If it is the former, then it means poloidal fields have been dragged into
the outer disk and can play a role in both the disk evolution and the
disk-wind launching \citep{Konigl2000}.  If it is the latter, then it means
the grains have grown from $\sim$ 0.1 \micron{} in the ISM to 100 \micron{}
in the outer disk, and this grain growth can facilitate an earlier start of
planet formation in the earliest phase of star formation.  Since the dust
disk appears to be geometrically thick, the grain settling is not
significant, possibly because of turbulence produced by an active accretion. 
Recent multi-wavelength continuum observations have shown that the disk is
likely subject to gravitationally instability \citep{Tobin2020},
supporting that the disk is in an active accretion phase.



Recently, dust polarization due to aligned grains has also been detected on
the larger scales in the dense molecular core and the flattened envelope
around the disk, revealing magnetic field morphology there
\citep{Yen2021,Galametz2020}, and thus allowing us to check the possible
presence of poloidal fields in the outer disk.  In the molecular core at a
size scale of $\sim$ 0.1 pc, the magnetic fields are found to be poloidal
with a mean axis at a position angle of $\sim$ 35\degree{}$\pm10$\degree{},
slightly misaligned with the disk axis (of symmetry) by $\sim$
12\degree{} counterclockwise \cite[see Figure 11 in][]{Yen2021}.  Note that
the disk axis has a position angle of $\sim$ 23\degree{}, well aligned with
the jet axis \citep{Lee2017Disk}.  In the flattened envelope at a size scale
of $\sim$ 1500 au, the magnetic fields are found to have a similar mean axis
to that in the molecular core \citep{Galametz2020}.  Interestingly, the
innermost flattened envelope with a size scale of $\sim$ 500 au detected in
dust continuum also has an axis with a similar position angle of $\sim$
36\degree{}$\pm10$\degree{} \cite[see Figure 1c][]{Lee2017Disk}.  This
supports the notion that magnetic fields play an important role in the
collapsing process of the molecular core and that the flattened envelope is
a pseudo-disk formed by magnetically guided collapse \citep{Hirano2019}. 
Similar pseudo-disks have also been detected in HH 211 \citep{Lee2019HH211}
and OMC-3/MMS6 \citep{Liu2020}.  In this scenario, the poloidal fields are
dragged into the flattened envelope from the molecular core.  These poloidal
fields can then be dragged into the outer disk.  Recent detection of disk
wind from the outer disk
\citep{Tabone2017,Lee2018Dwind,Tabone2020,Lee2021Dwind} also suggests a need
for poloidal fields there.

The contribution of dust self-scattering to the continuum emission of this
disk has also been briefly investigated recently.  By modeling the continuum
emission maps in 3 wavelengths (from 0.852 to 3 mm) simultaneously,
\citet{Lin2021} found that a pure thermal dust emission can roughly
reproduce the dust continuum in those wavelengths.  This suggests that the
dust self-scattering, even if present at submillimeter wavelength, could be
relatively small in this disk.  Similar conclusion was obtained in
\citet{Galvan2018}.  Considering a maximum grain size of 60-150 \micron{}
estimated from scattering-induced polarization in Class I and II
protostellar systems in the more evolved phases
\citep{Kataoka2016a,Yang2016a,Bacciotti2018,Hull2018}, it is also reasonable
to consider a maximum grain size smaller than 100 \micron{} in HH 212 in
the Class 0 phase.

In this paper, we present our new dust polarization detection in the HH 212
disk in continuum, obtained with Atacama Large Millimemeter/submillimeter
Array (ALMA) at 3 times higher resolution than previous observations, in
order to further determine the contributions of dichroic extinction and dust
self-scattering.  By modeling the polarization morphology, polarized
intensity, and polarization degree, we find that neither mechanisms can
fully reproduce the observations individually and that both mechanisms might
be required.  Being edge-on and vertically resolved, the HH 212 disk
provides the best view to search for poloidal fields and to study the grain
growth and settling that are the crucial first steps towards planet
formation.

\section{Observations}\label{sec:obs}

Linear dust polarization observations of the HH 212 disk were obtained
with ALMA in Band 7 in Cycle 5 (Project ID: 2017.1.00044.S).  Three
observations were taken on 2017 November 27, with a total time of $\sim$ 98
minutes on the target.  A single pointing was used to map the disk with a
primary beam of $\sim$ \arcs{17}.  47 antennas were used in the
observations, with projected baselines of $\sim$ 60$-$8500 m.  The maximum
recoverable size scale was $\sim$ \arcsa{0}{4}, enough to map the disk
without any significant missing flux.  The correlator was set up to have 4
spectral windows (centered at 334.7, 336.5, 346.6, and 348.4 GHz,
respectively), with a total bandwidth of $\sim$ 8 GHz centered at $\sim$
341.5 GHz (or 878 \micron{} correspondingly).

The Common Astronomy Software Applications (CASA) package was used to
calibrate the $uv$ data manually by the ALMA QA2 team, with quasar
J5010+1800 as a passband and flux calibrator, and quasar J0541-0211 as a
gain calibrator, and quasar J0522-3627 as a polarization calibrator.  We
adopted a super-uniform weighting with a robust factor of 0.5 (with npixels=0)
for the $uv$
data to generate the continuum map of the disk at $\sim$ 341.5 GHz with a
synthesized beam (resolution) of $\sim$
\arcsa{0}{036}$\times$\arcsa{0}{032}.  We also performed a phase-only
self-calibration using the continuum intensity (Stokes-I) map to improve the
map fidelity.  In Stokes \iI{} map, the noise level is $\sigma \sim$ 25
\uJyb{} (or 0.233 K).  In Stokes \iQ{} and \iU{} maps, the noise level is
$\sigma_p \sim$ 19 \uJyb{} (or 0.177 K).  From the Stokes parameters, we can
derive the linear polarized intensity, polarization fraction, and
polarization orientation.  Linear polarized intensity is defined as
$P_i=\sqrt{Q^2+U^2-\sigma_p^2}$ and thus bias-corrected.  Then polarization
fraction is defined as $P=P_i/I$.  According to ALMA Technical Handbook in
Cycle 5, the instrumental error on $P$ is expected to be $\lesssim$ 0.2\%
for the disk, which has a size much smaller than 1/3 of the primary beam. 
Polarization orientations are defined by the $E$ vectors.


\section{Results}


Throughout this paper, in order to facilitate our presentations, we rotate
our maps by 23\degree{} clockwise to align the major axis of the disk in the
horizontal direction.  Figure \ref{fig:HH212map} shows the maps from
our polarization observations towards the disk in continuum at $\lambda\sim$
878 \micron{}.  As seen before at $\lambda \sim$ 852 \micron{} in
\citet{Lee2017Disk}, the continuum intensity map shows a ``hamburger" like
emission structure for the disk, with a dark lane along the equatorial plane
sandwiched by two brighter features arising from the upper and lower disk
surfaces.  As discussed in that paper, the brighter feature above is
slightly brighter than the one below, because the nearside of the disk is
tilted slightly to the south.  In addition, the emission is mainly from the
outer edge of the disk on the nearside, because of nearly edge-on
orientation, large geometric thickness, and high optical depth of the disk.

Polarized emission has been detected before at a similar wavelength
\citep{Lee2018BDisk}.  Now at 3 times higher resolution, we can clearly see
that the polarized emission is mainly detected in the dark lane within the
centrifugal barrier, which has a radius of $\sim$ \arcsa{0}{11} (44 au),
where a rotationally supported disk has formed \citep{Lee2017COMs}. 
Interestingly, the polarized intensity shows two peaks in the midplane near
the central source on either side at a distance of $\sim$ 0.02 (8 au), with
a small dip at the central source (see also Figure \ref{fig:HH212mapdeg}). 
The western peak (right) is brighter than the eastern peak.  In addition,
the polarized emission extending to the east also peaks near the two (upper
and lower) disk surfaces, forming a V-shaped structure near the central
source (delineated by the white dotted curve in Figure \ref{fig:HH212map}b)
opening to the east, and then an arm structure further in the east near the
midplane.   The polarized emission extending to the west could also form
such a V-shaped structure near the central source opening to the west, but
observations at higher resolution are needed to resolve and confirm it. 
Nonetheless, it also shows an arm structure further in the west in the
midplane. Since the disk is subject to gravitational instability
\citep{Tobin2020} and thus might have spiral arms as seen in the HH 111 disk
\citep{Lee2020HH111}, the polarization emission near and in the midplane
could be affected by them, forming polarized arm structures there.

The polarization orientations (as indicated by the line segments) are rather
uniform and mostly parallel to the disk minor axis, which is the axis
perpendicular to the disk midplane.  Going away from the midplane to the
disk surfaces, the polarization orientations in the west are
tilted slightly away from the minor axis to form a
concave shape with respect to the center.
The polarization degree peaks at $\sim$ 3-4\% near
the center at the two polarized intensity peaks and decreases outwards along
the major and minor axes to $\sim$ 1.5\% at the edges and surfaces of the
disk (see also Figure \ref{fig:HH212mapdeg}).  The polarized intensity
further out drops below the 3$\sigma$ sensitivity, as shown in Figure
\ref{fig:HH212mapdeg}.



\section{Polarization Models} \label{sec:model}

In our previous study of dust polarization of this disk at similar
wavelength at a lower resolution of $\sim$ \arcsa{0}{12}
\citep{Lee2018BDisk}, the morphology of the polarized emission was not
resolved and the polarization could be due to either dichroic extinction by
dust grains aligned by poloidal magnetic fields or dust self-scattering by
large grains with a size as large as 100 \micron{}.  Following up on that,
we first introduce a simple parametrized disk model and then explore these
two possible mechanisms in more details.

\subsection{A Flared Dusty Disk Model} \label{sec:dmodel}


Similar flared disk models have been used to produce the thermal dust
emission in this disk \citep{Lee2017Disk,Galvan2018,Lin2021}.  Here we adopt
a similar model to reproduce the observed dust polarization.  The
disk is composed of dust and gas in vertical hydrostatic equilibrium with a scale height of $\hs$. 
In a cylindrical coordinate system, the disk is assumed to have the
following dust mass density and temperature
\begin{eqnarray} 
\rho (R,z) &=& \rhoa (\frac{R}{\Ra})^{-p} \exp(-\frac{z^2}{2 \hs^2}) \nonumber \\
T (R,z) &=& \Ta (\frac{R}{\Ra})^{-q} \exp(\frac{z^2}{2 \hs^2})
\end{eqnarray}
where $\Ra$ is a reference radius to be defined below, $\rhoa$ and $\Ta$ are
the dust mass density and temperature in the disk midplane at $\Ra$, respectively,
and $p$ and $q$ are the power-law indexes.  As in \citet{Lee2017Disk},
we assume that $p=2$ and $q=0.75$. 
Assuming a gas to dust mass ratio of $100$
and a gas composed of molecular hydrogen and atomic helium, we have the number
density of molecular hydrogen 
at $\Ra$ given by
\begin{equation}
\na = \frac{100\, \rhoa}{1.4\, \mHt}
\end{equation}
Here the number density of Helium is assumed to be $0.2$ times that of molecular hydrogen.

The scale height is assumed to be
proportional to $\frac{\cs}{\vp} R$, where $c_s$ is the isothermal sound
speed proportional to $T^{1/2}$ and thus to $R^{-q/2}$, and $\vp$ is the
rotational velocity assumed to be Keplerian and thus proportional to
$R^{-1/2}$ \citep{Lee2017COMs}.  As a result, the scale height is given by

\begin{equation}
\hs (R) = \ha (\frac{R}{\Ra})^{1+(1-q)/2}
\end{equation}
with $\ha$ being the scale height at $\Ra$. The total height of the disk can reach
$\sqrt{2} \hs$.  However, as discussed in \citet{Lee2017Disk}, since the
continuum emission of the disk becomes geometrically thinner near the outer
edge, the total height of the disk is revised to 
\begin{eqnarray}
\ho (R)= \sqrt{2}\hs \left\{ \begin{array}{ll} 1 & \;\;\textrm{if}\;\; R < \Ra, \\
\exp[-(\frac{R-\Ra}{\Ro-\Ra})^2] & \;\;\textrm{if}\;\; \Ra \leq R \leq \Ro
\end{array} \right.  
\label{eq:thick} 
\end{eqnarray} 
where $\Ro$ is the outer radius of the disk and $\Ra$ is
the radius beyond which the total height of the disk is tapered
to roughly match the observed height in the outer edge. 
Then the disk
has a surface density of dust given by
\begin{equation}
\Sigma(R) = \int_{-\ho}^{\ho} \rho \,dz 
\end{equation} 



Radiative transfer assuming LTE is used to calculate the dust emission from
the model, using our radiative transfer code in \citet{Lee2017Disk}. 
A major uncertainty in the model is the dust absorption opacity.  Recently,
multiple-wavelength observations in longer wavelengths have been used to
constrain it.  Using the dust continuum emission at 0.87 and 9 mm at $\sim$
\arcsa{0}{1} resolution, \citet{Tobin2020} has found that the disk has
Toomre Q values of 1$-$2.5.  By modeling the continuum emission maps in 3
wavelengths (from 0.852 to 3 mm, including a wavelength similar to the
observed here) simultaneously at higher resolution assuming pure thermal
dust emission, \citet{Lin2021} have estimated the absorption opacity to be
$\kabs \sim 1.9\, Q$ cm$^2$ per gram of dust at the wavelength of $\sim$ 852
\micron{}.  Thus here in our model, adopting $Q \sim 1$, we assume $\kabs
\sim 1.9$ cm$^2$ per gram of dust at the observed wavelength of 878
\micron{}.  Dust self-scattering opacity will be added later when we study
the contribution of dust self-scattering to the dust polarization.

The disk is assumed to be nearly edge-on with an inclination of $\sim$
87\degree{} \citep{Lee2021Dwind} and the nearside tilted slightly to the south. 
By matching the observed structure of the continuum emission assuming pure
dust thermal emission, we find that $\Ra \sim$ 34$\pm5$ au (or
\arcsa{0}{085}$\pm$\arcsa{0}{013}), $\Ro \sim 68\pm10$ au (or
\arcsa{0}{17}$\pm$\arcsa{0}{03}), $\hs \sim$ 12$\pm2$ au (or
\arcsa{0}{03}$\pm$\arcsa{0}{005}), $\rhoa \sim 3.0\pm0.4 \times 10^{-15}$ g
\cmc{} (or $\na \sim 6.5\pm1.0\times10^{10}$ \cmc{}), and $\Ta \sim 65\pm10$
K, consistent with those found in \citet{Lee2017Disk}, but with a factor of
3 higher $\na$ due to a factor of 3 lower absorption opacity.  Thus, the
disk has a total (gas plus dust) mass of $\sim 0.14\pm0.02$ \solarmass{},
which is also about a factor of 3 higher than that derived in
\citet{Lee2017Disk} and thus becomes $\sim$ 56\% of the protostellar mass. 
This mass is also consistent with that found by \citet{Galvan2018}, who
assumed a similar absorption opacity at 878 \micron{} according to their
adopted opacity law. This disk is so massive and thus subject to
gravitational instability, as discussed in \citet{Tobin2020}.




\subsection{Dichroic Extinction by Magnetically Aligned Grains} \label{sec:modelBp}

Dichroic extinction is the differential attenuation of two orthogonal
components of the $E$ vector of light by magnetically aligned grains
\citep{Wood1997}, resulting in a net polarization of light. In this
scenario, we assume that the disk is threaded with uniform (vertical)
poloidal magnetic fields and the dust grains are aligned by the magnetic
fields, with their long axis perpendicular to the field direction
\citep{Andersson2015}.  Because the bulk of the disk is optically thick at
our observing wavelength and the temperature increases along the line of
sight into the disk, dichroic extinction (rather than direct emission) by
aligned grains is expected to determine the degree and orientation of the
polarization relative to the magnetic field direction \cite[see, e.g.,
Figure 1 in][for an illustration]{Lin2020}.



\def\Cextb{\bar{C}_\textrm{\scriptsize ext}}
\def\Cext{C_\textrm{\scriptsize ext}}
\def\Cextx{C_\textrm{\scriptsize ext,x}}
\def\Cexty{C_\textrm{\scriptsize ext,y}}
\def\Cextpa{C_\textrm{\scriptsize ext,$\parallel$}}
\def\Cextpe{C_\textrm{\scriptsize ext,$\perp$}}
\def\Cpol{C_\textrm{\scriptsize pol}}
\def\Cpolb{\bar{C}_\textrm{\scriptsize pol}}
\def\CpolR{C_\textrm{\scriptsize pol}}  
\def\Cpolq{C_\textrm{\scriptsize pol,q}}
\def\Cpolu{C_\textrm{\scriptsize pol,u}}
\def\Ccircb{\bar{C}_\textrm{\scriptsize circ}}
\def\Ccirc{C_\textrm{\scriptsize circ}}
\def\Cabs{C_\textrm{\scriptsize abs}}
\def\Cabsb{\bar{C}_\textrm{\scriptsize abs}}
\def\Cabsq{C_\textrm{\scriptsize abs,q}}
\def\Cabsu{C_\textrm{\scriptsize abs,u}}
\def\Cran{C_\textrm{\scriptsize ran}}

Our radiative transfer code can be expanded to
calculate the polarization maps. With an assumption of LTE, 
the radiative transfer in
Stokes $I$, $Q$, and $U$ parameters can be given by the following
\cite[see also POLARIS in][]{Reissl2016}
\begin{eqnarray}
\frac{d}{\rho ds}  
\left( \begin{array}{c}
I \\
Q \\
U 
\end{array} \right) \;
= -
\left( \begin{array}{ccc}
\Cext & \CpolR \cos 2 \psi & \CpolR \sin 2 \psi  \\
\CpolR \cos 2 \psi & \Cext & 0  \\
\CpolR \sin 2 \psi &  0    & \Cext
\end{array} \right)
\left( \begin{array}{c}
I\\
Q \\
U 
\end{array} \right)
+ B_\nu(T) 
\left( \begin{array}{c}
\Cext\\
\CpolR \cos 2 \psi\\
\CpolR \sin 2 \psi
\end{array} \right)
\label{eq:pRTM}
\end{eqnarray}
The extinction cross section per gram (i.e., extinction opacity)
and polarization cross section per gram (i.e., polarization opacity)
can be given by \cite[see][]{Lee1985}
\begin{eqnarray}
\Cext &=& \kabs [1-\alpha(\frac{\cos^2 \gamma}{2} - \frac{1}{3})] \nonumber \\
\CpolR &=& \alpha \kabs \cos^2 \gamma
\end{eqnarray}
where $\gamma$ is the angle between the local $\mathbf{B}$ vector and the
plane of the sky, and $\psi= \phi + 90^\circ$ is the polarization angle measuring
from the north (ordinate axis) to the east, with $\phi$ being the angle
between the projection of the local $\mathbf{B}$ vector on the plane of the sky and north. 
In order to calculate the polarization quantities, we assume a
polarization efficiency $\alpha$, which defines the maximum polarization
fraction in an optically thin region \citep{Fiege2000,Padoan2001}. 
After the integration along each line of sight,
the resulting polarization intensity, fraction, and angle can then be given respectively by
\begin{equation}
P_i=\sqrt{Q^2+U^2}
\end{equation}
\begin{equation}
P=\frac{P_i}{I}
\end{equation}
and
\begin{equation}
\psi = \frac{1}{2} \tan^{-1} \frac{U}{Q}
\end{equation}
By matching the polarization fraction of $\sim$ 0.03 observed in the optically
thick central region of the disk, we find that a value of $\alpha \sim
0.075$ is required in the model. In this case, $\Cext \approx \kabs$.

Figures \ref{fig:HH212Pol} shows the model results.  The total intensity map
of the emission shows a dark lane along the major axis sandwiched by two
brighter curved features above and below, because the midplane of the disk
is cooler and optically thicker than the surfaces.  Since the disk's
nearside is tilted slightly to the south, the upper disk surface is slightly
exposed and thus becomes brighter than the lower one.  The whole disk is
optically thick.  In particular, the optical depth peaks at the center with
a value greater than 100, and it decreases outward and drops to $\sim$ 1
near the outer edges and outer disk surfaces.  The emission in the disk is
polarized.  Along the major axis, the polarized intensity peaks at the
center and decreases outward and drops to zero at $\sim$ \arcsa{0}{11} (44
au) where the optical depth drops to 3$-$5, and then increases again in the
optically thinner outer edges.  The initial decrease of the polarization
fraction with distance is physically reasonable since the fraction of
polarization produced by dichroic extinction along high optical depth
sight-lines is determined mainly by the temperature gradient along the line
of sight (los), which is the highest toward the center.  As the sightline
moves towards the outer edge of the disk, both the optical depth and the
(los) temperature gradient decrease, weakening the contribution of the
dichroic extinction to the polarization relative to that of the dichroic
emission.  Going away from the midplane to the disk surfaces, the polarized
intensity increases with the increasing distance and peaks near the disk
surfaces where the temperature is higher, outlining the disk surfaces.  Then
the polarized intensity drops to zero at $\sim$ \arcsa{0}{03} (12 au) where
the optical depth drops to 3$-$5, and then increases again in the optically
thinner outer surfaces.  As a result, when the optical depth drops to 3$-$5,
the polarized intensity and the resulting polarization degree drop to zero,
forming an elliptical polarization gap around the inner disk.  Interior to
the polarization gap in the inner optically thicker region, the polarization
orientations are parallel to the poloidal field direction and thus the minor
axis, because of the  dichroic extinction and the temperature gradient,
as discussed in \citet{Lee2018BDisk} and \citet{Lin2020}. 
 Going across the polarization gap to the outer optically
thinner region, the polarization orientations flip by 90\degree{} to be
perpendicular to the poloidal field (and thus the minor axis) direction. 
In this outer region, direct emission by aligned grains is dominant.


Figure \ref{fig:HH212Polc} shows the model results convolved to the observed
beam to be compared with the observations.  The total intensity map of the
emission also shows a dark lane sandwiched by two brighter features, similar
to that seen in the observations.  The polarization gap can still be seen in
the convolved map around the inner disk.  Interior to the polarization gap,
the polarized emission from the upper and lower disk surfaces merge and form
a single elongated polarization structure in the dark lane, with hints of
V-shaped structures opening to the east and west still discernible, although
far less prominent compared to the unconvolved map.  This polarization
structure, although less clearly V-shaped, could correspond to that observed
in the dark lane, which shows  a V-shaped structure opening to the east
and probably another one opening to the west.  Note that we do not intend
to reproduce the polarized arm structures, which could be affected by
potential spiral arms, as discussed earlier.  Both the polarized intensity
and polarization degree of the beam convolved model peak at the center and
decrease outward, also roughly similar to the observed trends.  The
polarization orientations are parallel to the minor axis, also similar to
the observed orientations, except for those in the west near the disk
surfaces where the observed orientations are tilted 
slightly away from the minor axis to form a concave shape with respect to the center.
However, unlike the observations that show two polarized emission
peaks near the central source  with a small dip in between, only one
single polarized emission peak is seen in the model toward the central
source.
  
In the model, the polarized intensity drops to zero in the polarization gap
and then increases outside.  Outside the polarization gap, although the
polarized emission is much fainter than that inside, the polarized emission
in the upper surface and outer disk above the 3$\sigma$ sensitivity
level (encompassed by the orange contour) in Figure \ref{fig:HH212Polc}b
should still be detectable in the current observations, with a flip of the
polarization orientation to be perpendicular to the minor axis.  However,
such an increase in polarized intensity and a flip of polarization
orientation in the outer disk are not detected here.  Thus, the polarization
gap and the increase of polarized intensity around the disk predicted in
this model might not exist in the observations but deeper observations are
needed to check them.

\subsection{Dust Self-Scattering by Submillimeter-Sized Grains}


Dust self-scattering by large grains (100 \micron{} or larger) has been
found to contribute significantly to the (sub)millimeter dust polarization
in the protoplanetary disks in the later phase of star formation
\citep{Kataoka2015,Kataoka2016a,Yang2016a,Yang2016b,Yang2017}, as seen in
the evolved disks in, e.g., HL Tau \citep{Kataoka2017,Stephens2017}, CW Tau
and DG Tau \citep{Bacciotti2018}, HD 142527 \citep{Ohashi2018}, IM Lup
\citep{Hull2018}, and HD163296 \citep{Dent2019}.  It can also contribute to
the dust polarization in the early phase in the protostellar disks if the
dust grains have grown to 100 \micron{} in size
\citep{Sadavoy2018a,Alves2018,Harris2018,Sadavoy2018b}.  Since the
scattering opacity is still uncertain, we parametrize it by
\begin{equation} \ksca = \frac{\omega}{1-\omega} \kabs \end{equation} where
$\kabs$ is the absorption opacity and $\omega$ is the albedo defined as
\begin{equation} \omega = \frac{\ksca}{\kabs+\ksca} \end{equation} This
albedo can be linked to grain size.  For example, with a size distribution
of $n(a)\propto a^{-3.5}$ and a mixture of silicate, water ice, and
organics, a maximum grain size of 10, 25, 50, 75, 100 \micron{},
\citet{Kataoka2015} estimated an 
albedo of $\sim$ 0, 0.036 0.23, 0.50, 0.70, respectively, at the observed
wavelength of 870 \micron{}, 
although the exact values depend on grain properties (such as shape and
composition), which are uncertain.

 Since the disk here is young in the  deeply embedded, actively accreting
protostellar phase, the grain size is expected to be smaller than 100
\micron{} \citep{Galvan2018,Lin2021}.  Since the dust polarization
morphology due to dust self-scattering does not depend much on the actual
size of the grains (as long as the grains are not too large compared to the
observing wavelength) and the grain composition, we adopt an albedo of 0.5,
or equivalently assume a scattering opacity the same as the absorption
opacity, to study the dust self-scattering effect.  For demonstration
purpose, we first derive the dust scattering matrix with the Mie
approximation using the optical constants of an amorphous silicate
(amorphous pyroxene with 70 percent magnesium and 30 percent iron)
\citep{Jaeger1994,Dorschner1995}, a grain size of 100 \micron{}, and
material density of 1.675 g \cmc{} \citep{Birnstiel2018}, and then scale it
down to have a scattering opacity the same as that of the absorption opacity
to be consistent with our adopted albedo.  Then we use the
RADMC-3D\footnote{RADMC-3D is a publicly available code for radiative
calculations available at
http://www.ita.uni-heidelberg.de/$\sim$dullemond/software/radmc-3d/} code
\citep{Dullemond2012} to produce the polarization maps.  Since the
dust-scattering opacity adds an additional optical depth and thus an
attenuation of the dust emission, a higher temperature with $\Ta \sim
95\pm14$ K, which is about 50\% higher than that without dust scattering, is
required to roughly match the observed brightness temperature.





Figure \ref{fig:HH212Scat} shows the model results.  Like the dichroic
extinction model, the total intensity map also shows a dark lane in the
major axis sandwiched by two brighter features arising from the disk
surfaces.  The dark lane is wider because of a larger optical depth and the
emission is scattered away from the midplane.  However, unlike the dichroic
extinction model, no polarization gap is seen.  In addition, the polarized
intensity shows a dip at the center because the radiation is most symmetric
about the center.  It increases outwards in all directions and peaks near
the edges and at the disk surfaces, because of the increase of asymmetry and
also the increase of temperature for the disk surfaces.  It then drops
rapidly to zero at the edges and above the surfaces.  Like the total
intensity, the polarized intensity from the upper disk surface is brighter
than that from the lower one.  The polarization degree also shows a similar
trend to the polarized intensity, except near the two edges where the
polarization degree continues to increase towards the edges due to the
faster decrease of total intensity.  The polarization orientations are
parallel to the minor axis near the midplane and then tilted slightly away
from the minor axis to form a concave shape with respect to the center
 when going away from the midplane to the upper and lower disk surfaces.

Figure \ref{fig:HH212Scatc} shows the model results convolved to the
observed beam.  The total intensity map shows a similar structure to that
observed.  The polarization orientations are mostly parallel to the minor
axis, and tilted slightly away from the minor axis 
to form a concave shape with respect to the center when going away from the
midplane to the disk surfaces, similar to that seen in the observations. 
Unlike the dichroic extinction model, the upper disk and lower disk surfaces
in the polarization intensity do not merge to form a single elongated
structure, because the dark lane is wider and the lower disk surface is much
fainter.  In addition, unlike those seen in the observations, the
polarized intensity and polarization degree in the model increase toward the
edges along the major axis and drop rapidly at the edges.  With the
polarized intensity peaks near the two edges, the resulting polarized
intensity dip is much wider than the observed, inconsistent with the
observations.  We have tried with a smaller albedo of 0.25.  In this case,
the polarized emissions from the upper disk and lower disk surfaces move
slightly closer to the midplane, forming V-shaped structures opening to the
east and west, roughly similar to the observed.  However, the polarized
intensity still peaks at the two edges, inconsistent with the observations. 
Moreover, the polarization intensity becomes everywhere lower than the
3$\sigma$ sensitivity limit in our current observations.


\section{Discussion}

Dust polarization is detected in the HH 212 disk within the centrifugal
barrier at $\sim$ 44 au, where a rotationally supported disk  has
formed.  The comparison of our simple models with the observations suggests
that neither a simple dichroic extinction model by magnetically aligned
grains nor a simple dust self-scattering model can fully account for the
observed polarization morphology, polarized intensity, and polarization
degree simultaneously.  The dichroic extinction model can produce an
elongated polarized structure in the dark lane, with hints of V-shaped
structures opening to the east and west, similar to the observations.
It can also produce the decrease of polarized intensity and polarization
degree in the dark lane from the center towards the edges.  However, it
can not produce a small polarized intensity dip at the center.  In addition,
it produces a polarization gap and polarized emission with polarization
orientations perpendicular to the minor axis outside the gap in the outer
disk, which are not seen in the observations.  Also, it can not produce the
observed slight tilt of polarization orientations away from the minor axis
in the western part of the disk above and below the disk midplane.  On the
other hand, the dust self-scattering model can roughly produce the observed
polarization orientations.  However, it can not produce the observed
decrease of polarized intensity and polarization degree from the center to
the outer edges.  Moreover, the polarized intensity in the model always
peaks near the two edges, producing a dip much wider than the observed. 
Adding a flattened envelope around the disk may reduce the polarized
intensity near the edges by shining radiation into the disk, but may not be
sufficient to reduce it down to zero.



None of our two simple models can reproduce the small observed dip of
polarized intensity towards the central source position.  This polarized
intensity dip is unlikely to be a polarization hole due to depolarization of
unresolved complicated polarization orientations, as seen in, e.g., Serpens
SMM1 \citep{Hull2017}, because the polarization orientations there are
rather uniform.  Our dichroic extinction model assumes a constant
polarization efficiency and thus grain alignment efficiency.  A decrease of
grain alignment efficiency toward the center, as proposed to explain the
decrease of polarization fraction toward the region with higher column
density \citep{Galametz2018}, may help to produce a small polarized
intensity dip at the center.  However, further work is needed to study this
possibility.  On the other hand, our self-scattering model assumes a single
albedo and thus the same maximum grain size all over the disk.  It is
possible that the maximum grain size may decrease from the inner to the
outer disk, as expected if the grain size increases across the centrifugal
barrier from the outer disk to the inner disk.  For example, the grains
newly accreted from the envelope to the outer disk may be too small to
scatter (sub)millimeter photons efficiently.  Further modeling is needed to
check if the decrease of maximum grain size toward the edges can produce a
small polarized intensity dip at the center and a decrease of polarized
intensity towards the edges.



It is also possible that both mechanisms contribute to the dust polarization
in the HH 212 disk.  The dichroic extinction model better produces the
decrease of polarized intensity and polarization degree in the dark lane
towards the edges, while scattering model better produces the tilt of
the polarization orientations away from the minor axis when going away from
the midplane to the disk surfaces.  In addition, combining these two models
may also produce two polarized intensity peaks near the central source with
a small polarization dip at the central source, as seen in the observation. 
Since RADMC-3D cannot treat scattering by aligned (non-spherical) grains at
the present time, we will defer the combined modeling to a future
publication.  As mentioned in the introduction, poloidal fields can be
dragged from the dense molecular core into the innermost part of the
envelope and then to the outer disk.  If this is the case, the magnetic
fields may be responsible for launching the rotating SO/SO$_2$ outflow
extending out from the disk \citep{Tabone2017,Lee2018Dwind,Lee2021Dwind}. 
Observations at higher resolution are needed to determine the location of
the magnetic fields more accurately.  Also as mentioned in the introduction,
recent multi-wavelength observations suggested that the dust self-scattering
contribution to the dust continuum emission should be smaller than the
thermal dust emission \citep{Lin2021}, otherwise the required temperature in
the disk would be too high.  As seen in our model with an albedo $\omega =
0.5$, the temperature in the disk is already $\sim$ 50\% higher than that
without dust scattering.  According to \citet{Kataoka2015}, albedos are
0.23, 0.50, and 0.70 for maximum grain sizes of 50, 75, and 100 \micron{},
respectively, which then result in $\ksca=0.23$, 1.0, and 2.33 $\kabs{}$,
respectively.  Thus, in order to have a scattering opacity comparable to or
smaller than the absorption opacity, the grains can have a maximum size of
50$-$75 \micron{}, which is reasonable for a Class 0 disk in the outer
edges.  Moreover, since the disk is vertically extended, the grain settling
is unlikely to have taken place significantly.  Hence, the grains are likely
relatively small.  Furthermore, smaller grains are better for magnetic
alignment.



We can estimate whether the midplane temperature inferred from our modeling
can be produced by viscous heating due to disk accretion.
Since the disk here is optically thick, the midplane temperature due to viscous
heating would be \citep{Shakura1973}
\begin{equation}
T_{\mid}(R) \sim \tau_\Ross^{1/4} T_{\eff} (R)
\end{equation}
where $T_{\eff}$ is the 
effective (surface) temperature of the disk at $\tau \sim 1$ given by
\begin{equation}
T_{\eff}(R) \sim \Big(\frac{3 G M \dot{M} }{8\pi\sigma R^3}\Big)^{1/4}
\end{equation}
and $\tau_{\Ross}$ is the Rosseland mean optical depth defined as
\begin{equation}
\tau_{\Ross} \equiv \frac{1}{2} \Sigma \kappa_{\Ross}
\end{equation}
with $\sigma$ here being the Stefan Boltzmann constant,
$\kappa_{\Ross}$ being the Rosseland mean opacity per gram of dust,
and $\Sigma$ being the surface density of dust.
In HH 212, with $M \sim 0.25$ \solarmass{} \citep{Codella2014,Lee2014,Lee2017COMs}
and $\dot{M} \sim 5\times10^{-6}$
\solarmass{} yr$^{-1}$ \citep{Lee2014}, we have $T_{\eff} \sim 20$ K at $\Ra$ or 34 au.
In our model, the dust surface density $\Sigma \sim 1.14$ g \cms{} at $\Ra$.
For a
disk temperature between 20 and 100 K at that radius, the black-body radiation peaks at wavelength
$\lambda_{peak} \sim 29 - 145$ \micron{}, according to the Wien's displacement
law.  Then, assuming a maximum grain size of 100 \micron{}
and judging from Figure 1 in \citet{Kataoka2015} that shows the dust opacity versus wavelength, we have $\kappa_\Ross <
200$ cm$^2$ g$^{-1}$ and thus $\tau_\Ross < 114$. 
Thus, the midplane temperature in the disk should be less than 65 K at
$\Ra$.
Additional heating can come from irradiation by the central stellar object,
so the midplane temperature of about 65~K at $\Ra$ inferred in our dichroic
extinction model appears reasonable.  We refrain from a detailed modeling of
the temperature structure because it depends on the dust opacities, which
are uncertain.  Nevertheless,
 a midplane temperature of $\sim$ 95 K at $\Ra$ as in our dust self-scattering
model would require a Rosseland mean opacity 
of $\sim$ 5 times higher and is thus too high.

\section{Conclusions}

We have resolved the dust polarization in the HH 212 disk in continuum at
$\lambda \sim 878$ \micron{} and found it to be mainly arisen within the
centrifugal barrier at $\sim$ 44 au, where a rotationally supported disk
has formed.  The polarized intensity forms V-shaped structures opening
to the east and probably west arising from the disk surfaces and arm
structures further away in the east and west.  The polarization orientations
are mainly parallel to the minor axis, with some in the western part
tilting slightly away from the minor axis to form a
concave shape with respect to the center
when going away from the midplane to the
upper and lower disk surfaces.  The polarized intensity and polarization
degree both peak near the central source with a small dip at the central
source and decrease towards the disk outer edges.  The observed tilt of
the polarization orientations away from the minor axis is more consistent with
that expected from dust self-scattering by 50$-$75 \micron{} grains, but the
observed decreases of polarized intensity and polarization degree from the
center to the edges are more consistent with those expected from dichroic
extinction by grains aligned by poloidal fields.  The decreases of
polarized intensity and polarization degree toward the edges may also be
consistent with dust self-scattering if the grain size decreases toward the
edges, as expected if the grain size increases across the centrifugal
barrier from the outer edges. It is possible that both mechanisms are
needed to produce the observed dust polarization, suggesting the presence of
both poloidal fields and grain growth in the disk.  In addition, the
polarized arm structures could be due to potential spiral arms excited in
the outer disk as the disk has been found to be gravitational unstable.

\acknowledgements

We thank the anonymous referee for insightful comments. C.-F.L. 
thanks S.-Y.  Liu for fruitful discussions on dust polarization due to
magnetically aligned grains and H.  B.  Liu for fruitful discussions on
grain growth and dust opacity.  This paper makes use of the following ALMA
data: ADS/JAO.ALMA\# 2017.1.00044.S.  ALMA is a partnership of ESO
(representing its member states), NSF (USA) and NINS (Japan), together with
NRC (Canada), MoST and ASIAA (Taiwan), and KASI (Republic of Korea), in
cooperation with the Republic of Chile.  The Joint ALMA Observatory is
operated by ESO, AUI/NRAO and NAOJ.  C.-F.L.  acknowledges grants from the
Ministry of Science and Technology of Taiwan (MoST 107-2119-M- 001-040-MY3)
and the Academia Sinica (Investigator Award AS-IA-108-M01).  ZYL is
supported in part by NASA 80NSSC18K1095 and 80NSSC20K0533 and NSF
AST-1815784.  Z.  Lin acknowledges support by an ALMA SOS award and
Jefferson Graduate Fellowship at the University of Virginia.


\def\nat{Natur}

\begin{figure} [!hbp]
\centering
\putfig{0.7}{270}{f1.eps} 
\figcaption[]
{Polarization observations towards the HH 212 disk in continuum at $\lambda \sim$ 878 \micron{}
at $\sim$ \arcsa{0}{036}$\times$\arcsa{0}{032} resolution.
The blue and red arrows show the approaching and receding sides of the jet axis, respectively.
The asterisk marks the possible position
of the central source, which is assumed to have a coordinate
\ra{5}{43}{51}{4086} and \dec{-1}{2}{53}{154}, after
comparing the total intensity maps between the observations and models discussed in the text.
(a) Total intensity map. Contours start at 10 $\sigma$ with a step of 40 $\sigma$, where
$\sigma=0.256$ K.
(b) Polarized intensity map. Contours start at 3 $\sigma_p$ with a step of 2 $\sigma_p$, where
$\sigma_p=0.173$ K.
(c) Total intensity (contours) and polarized intensity (color) maps.
(d) Total intensity (contours), polarized intensity (color) maps
and polarization orientations (line segments, red for detections of $(2.5-3)\sigma_p$
and cyan for detections greater than 3$\sigma_p$).  
\label{fig:HH212map}}
\end{figure}

\begin{figure} [!hbp]
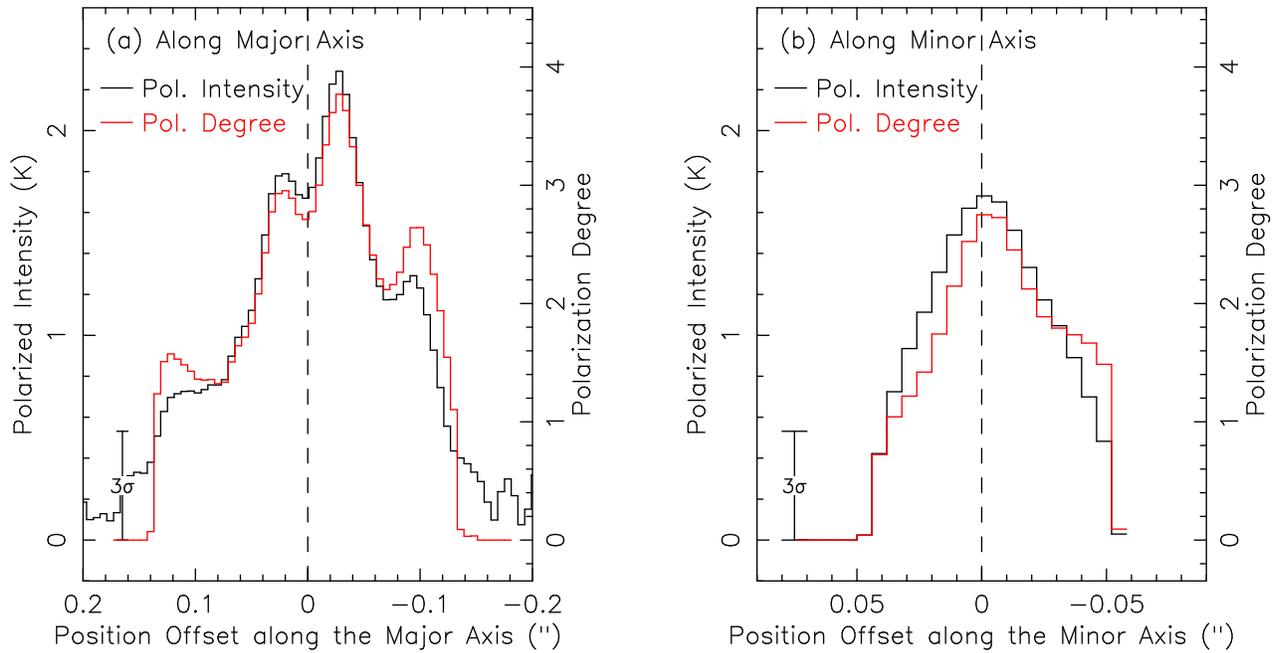

\centering
\putfig{0.7}{270}{f2.eps} 
\figcaption[]
{Polarized intensity (red curves) and polarization degree (black curves)
 at different position offsets from the central protostar
along the (a) major and (b) minor axes.
\label{fig:HH212mapdeg}}
\end{figure}

\begin{figure} [!hbp]
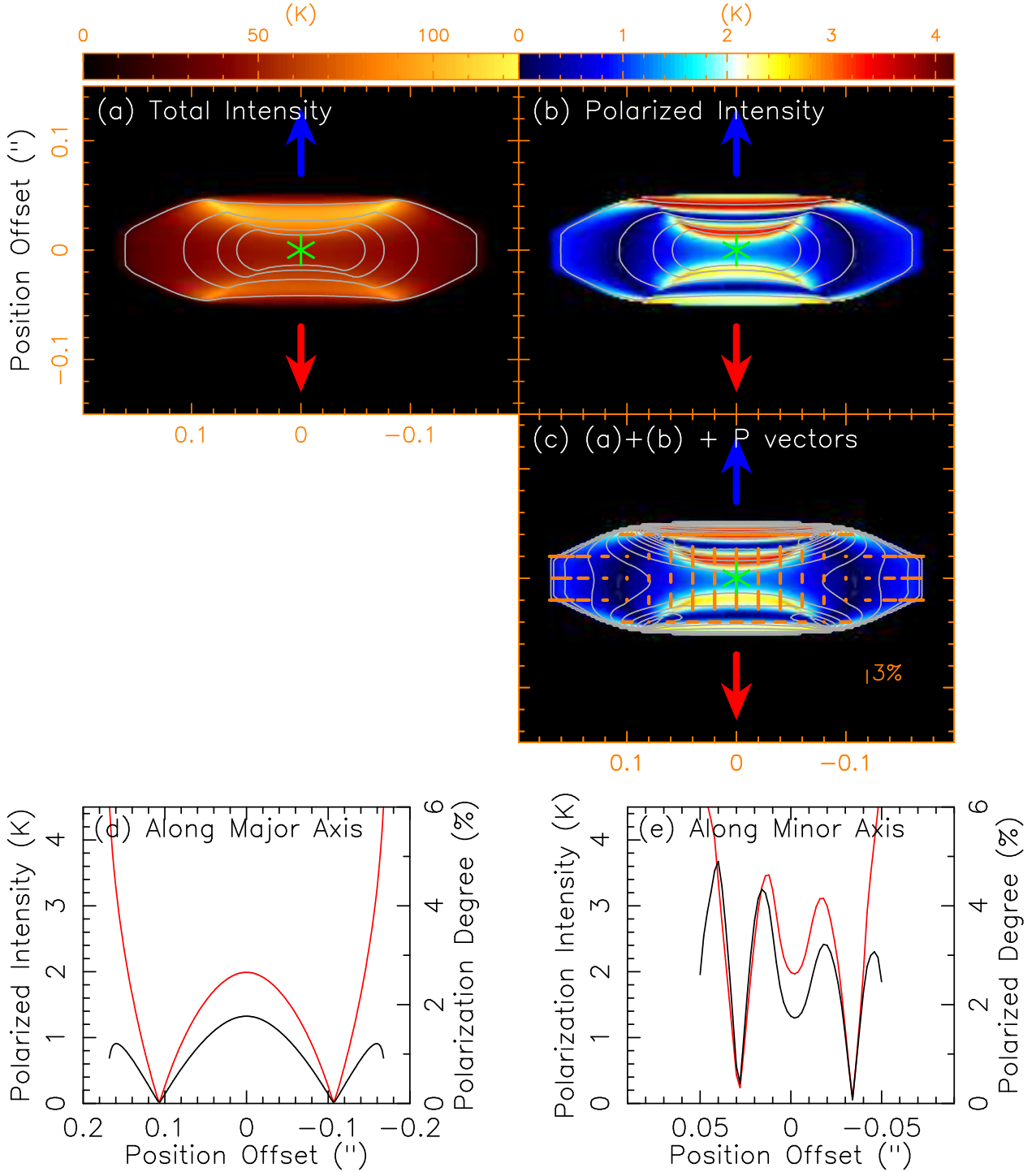

\centering
\putfig{0.9}{0}{f3.eps} 
\figcaption[]
{Dichroic extinction model by grains aligned magnetically by poloidal fields.
(a) Total intensity map (color) with optical depth (contours).
Contours start from 1 with a step of 3. Only first 4 contours are plotted.
(b) Polarized intensity map (color) with optical depth (contours).
(c) Total intensity map (contours), polarized intensity (color), and polarization degrees (line
segments with the length for the degrees).
(d) Polarized intensity (black curve) and polarization degree (red curve) along the major axis of the disk.
(e) Polarized intensity (black curve) and polarization degree (red curve) along the minor axis of the disk.
\label{fig:HH212Pol}}
\end{figure}

\begin{figure} [!hbp]
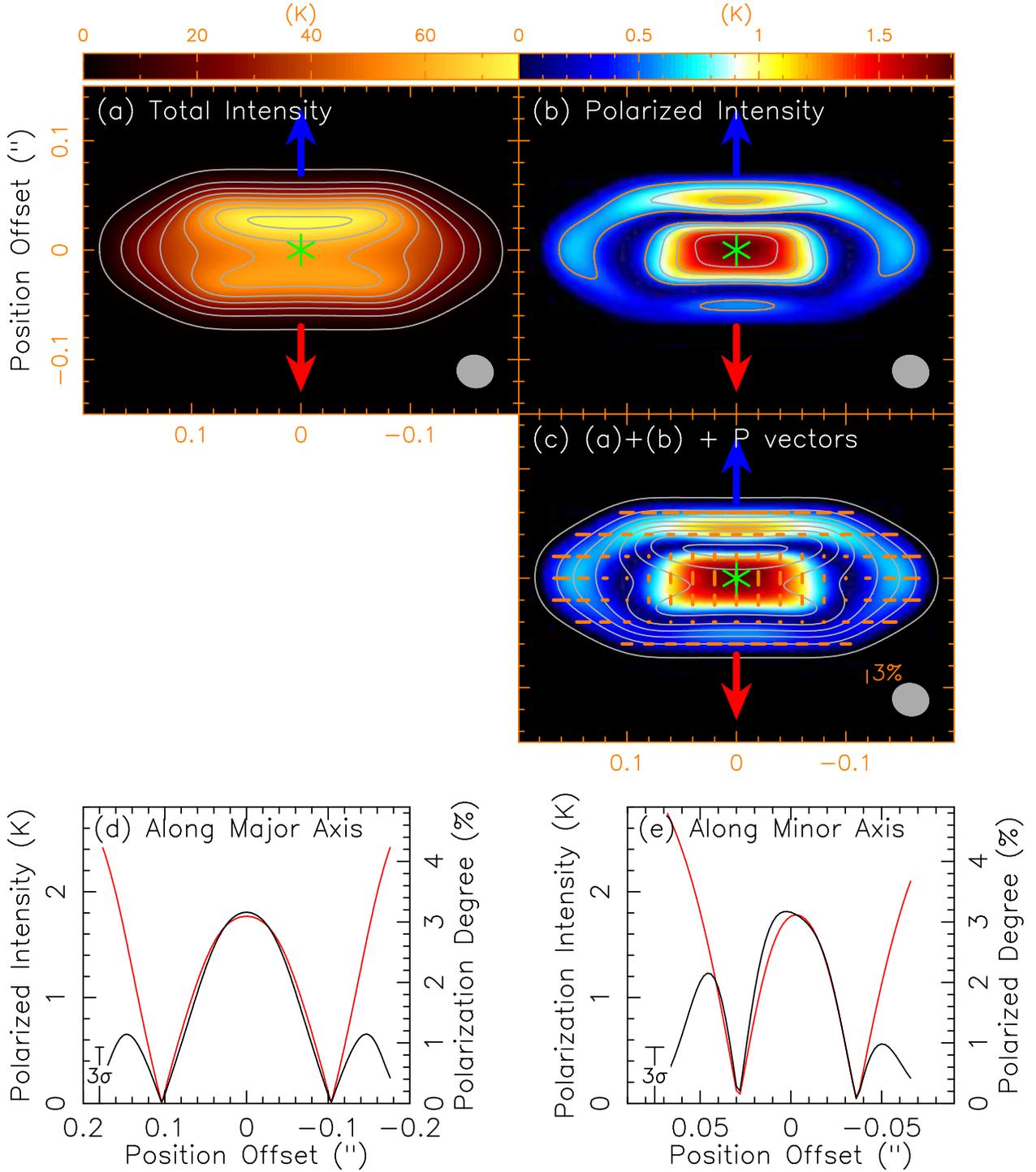

\centering
\putfig{0.9}{0}{f4.eps} 
\figcaption[]
{Same as Figure \ref{fig:HH212Pol} but convolved to the observed beam to be
compared with the observations.  The contour levels in (a) and (c)
are the same as those in Figure \ref{fig:HH212map}a. The contour levels
in panel (b) are the same as those in Figure \ref{fig:HH212map}b,
with the lowest contour (orange)
showing the 3 $\sigma$ sensitivity level in our observations.
\label{fig:HH212Polc}}
\end{figure}

\begin{figure} [!hbp]
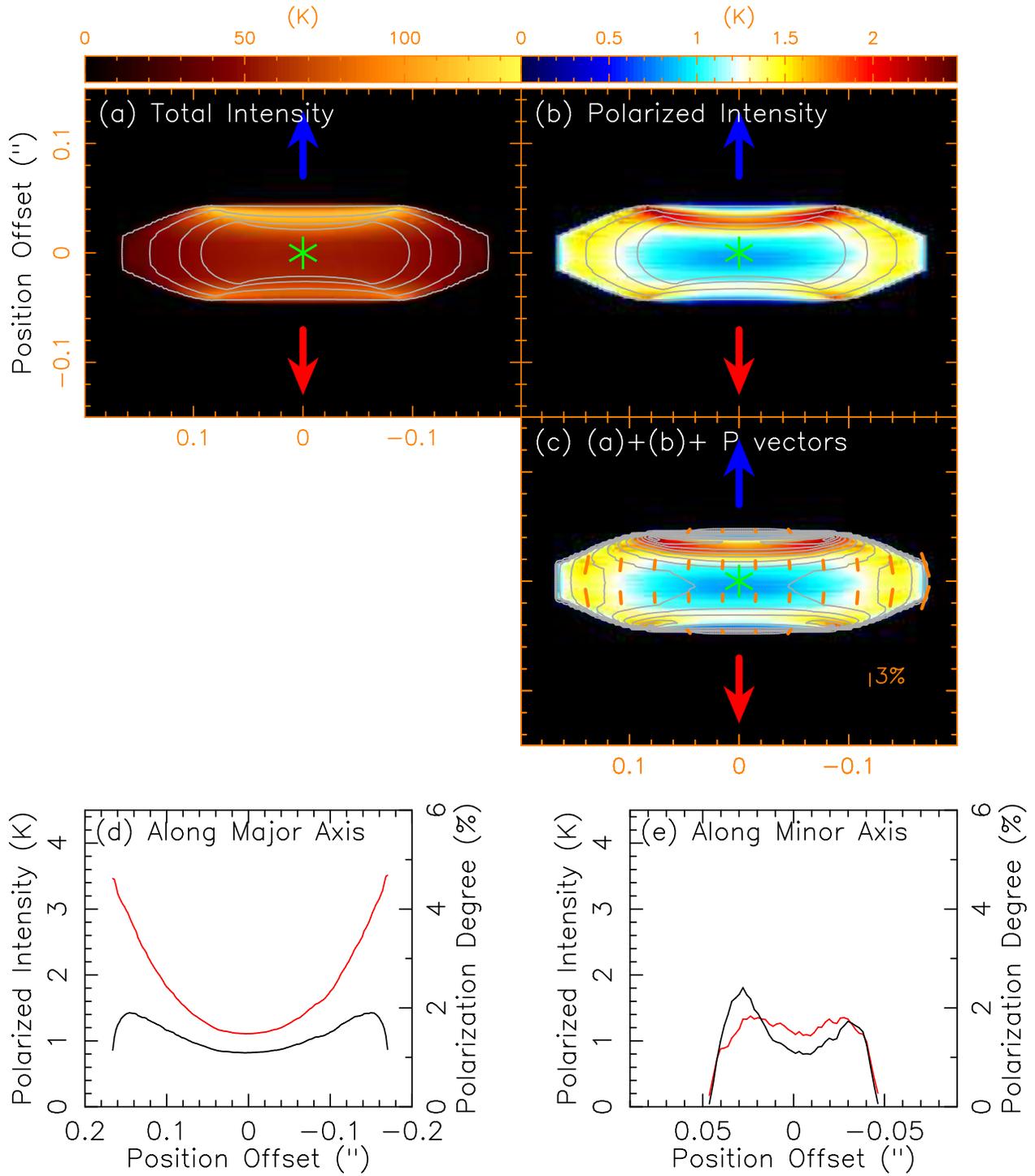

\centering
\putfig{0.9}{0}{f5.eps} 
\figcaption[]
{Dust self-scattering model. Same as Figure \ref{fig:HH212Pol}.
\label{fig:HH212Scat}}
\end{figure}

\begin{figure} [!hbp]
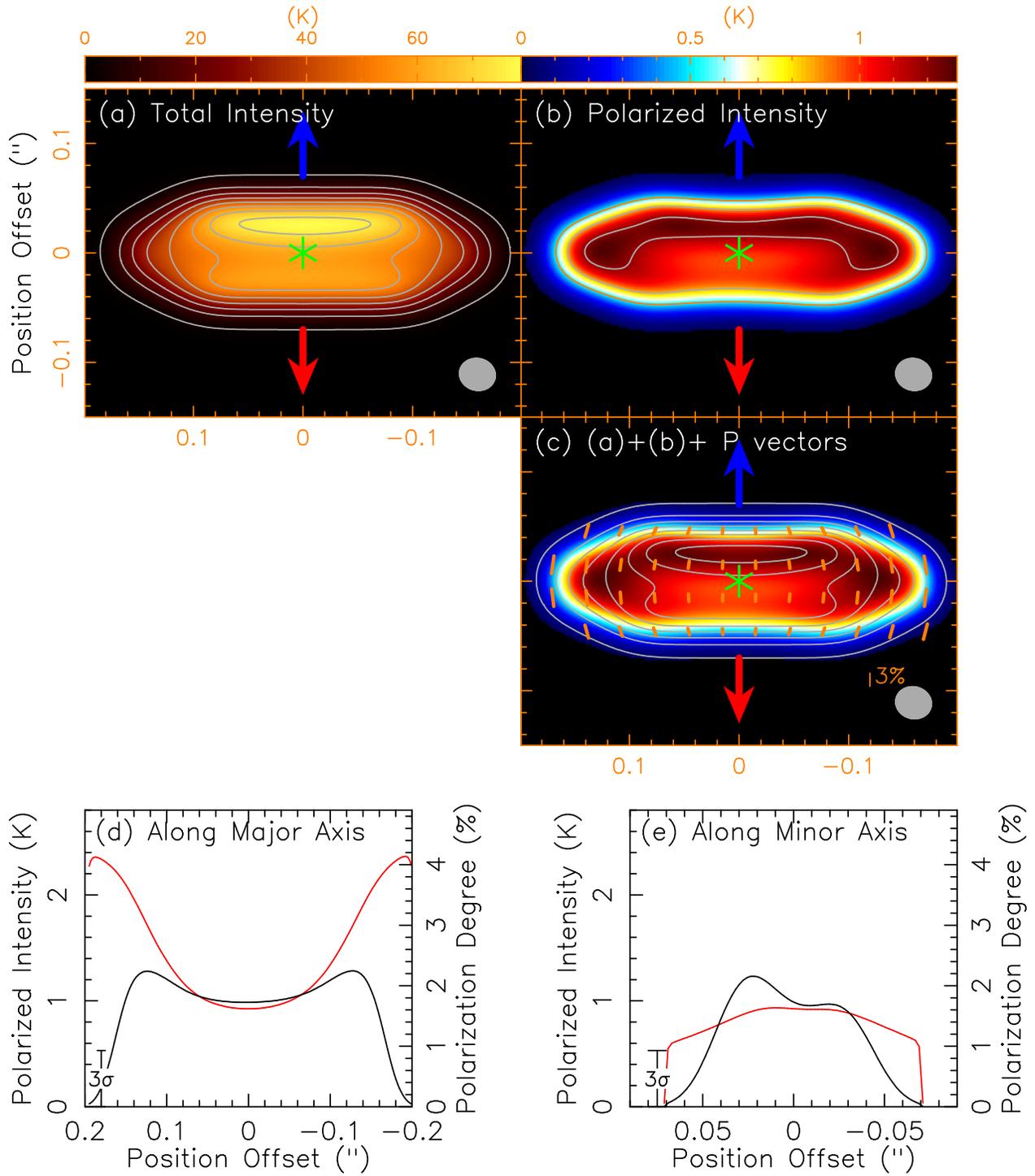

\centering
\putfig{0.9}{0}{f6.eps} 
\figcaption[]
{Same as Figure \ref{fig:HH212Scat} but convolved to the observed beam to be
compared with the observations.
The contour levels are the same as those in Figure \ref{fig:HH212Polc}.
\label{fig:HH212Scatc}}
\end{figure}


\end{document}